\documentclass[letterpaper]{JHEP3}
\title{Mollifying Quantum Field Theory or Lattice QFT in Minkowski Spacetime and
  Symmetry Breaking}
\author{D. D. Ferrante, G. S. Guralnik\\
  Department of Physics, High Energy Theory\\
  Brown University, Providence --- RI. 02912 \\
  E-mail: \email{danieldf@het.brown.edu}, \email{gerry@het.brown.edu}}
\preprint{BROWN-HET-1413\\ \heplat{0602013}}
\abstract{
  This work develops and applies the concept of mollification in order to smooth
  out highly oscillatory exponentials. This idea, known for quite a while in the
  mathematical community (mollifiers are a means to smooth distributions), is
  new to numerical Quantum Field Theory. It is potentially very useful
  for calculating phase transitions (highly oscillatory integrands in general),
  for computations with imaginary chemical potentials and Lattice QFT in
  Minkowski spacetime.
  }
\usepackage[T1]{fontenc}
\usepackage{ae,aecompl}
\usepackage{lmodern}
\usepackage{microtype}
\usepackage{ucs}
\usepackage[utf8x]{inputenc}
\usepackage{graphicx}
\usepackage{xspace}
\usepackage{amsmath,amsfonts,amstext,amssymb,amsbsy,amsopn,amsthm,eucal}
\DeclareMathAlphabet{\mathpzc}{T1}{pzc}{m}{it} 
\usepackage{mathrsfs}
\usepackage{wasysym}
\usepackage{empheq}
\usepackage{dsfont}
\usepackage{pifont}
\usepackage[boxed]{algorithm}
\usepackage{algorithmic}
\DeclareMathOperator{\Ai}{Ai}

\newcommand{\conv}[2]{\ensuremath{{#1}\ast{#2}}\xspace}

\newcommand{\cont}[1]{\ensuremath{{\mathscr{C}\/}^{#1}}\xspace}

\newcommand{\abs}[1]{\ensuremath{\left| #1\right|}\xspace}
\newtheorem{thm}{Theorem}[section]
\newtheorem{deff}[thm]{Definition}
\newtheorem*{notat}{Notation}
\def\hksqrt{\mathpalette\DHLhksqrt}
\def\DHLhksqrt#1#2{\setbox0=\hbox{$#1\sqrt{#2\,}$}\dimen0=\ht0
  \advance\dimen0-0.2\ht0
  \setbox2=\hbox{\vrule height\ht0 depth -\dimen0}%
{\box0\lower0.4pt\box2}}
\makeatletter
\def\equalsfill{$\m@th\mathord=\mkern-7mu
  \cleaders\hbox{$\!\mathord=\!$}\hfill
  \mkern-7mu\mathord=$}
\makeatother
\begin{document}
\section{Motivation and Introduction} \label{sec:intro}
\paragraph{Sign Problem}
One of the fundamental difficulties of Monte Carlo (MC) approaches is known as the ``sign
problem''. It is encountered when the functional integrals to be evaluated do not have a
positive definite measure. It is not related to any approximations or fundamental errors in
the MC scheme, but it describes the situation where the statistical error can become very
large. In general, any expectation value can be written as,

\begin{equation}
  \label{eq:genvev}
  \langle O \rangle = \frac{\varint O[\phi]\, \mu[\phi]}{\varint\mu[\phi]} \; ,
\end{equation}
where $\mu$ and $O$ are real-valued functions of the field variables and, in general, the
measure $\mu[\phi]$ need not be positive: in Lorentzian QFT this measure is
complex-valued and given by $e^{i\, S[\phi]}\, \mathcal{D}\phi$. However, if $\mu[\phi]$
changes sign, it cannot be considered a probability density (\textit{i.e.}, a measure).

The standard trick to avoid this problem is to modify the measure in the following way:
$\tilde\mu[\phi] = |\mu[\phi]|/\varint|\mu[\phi]|$. Then, one absorbs the sign of $\mu[\phi]$ in the
quantity to be measured:

\begin{equation}
  \label{eq:signtrick}
  \langle O \rangle = \frac{\varint O[\phi]\,\mathrm{sign}\bigl(\mu[\phi]\bigr)\,
    \tilde\mu[\phi]}{\varint\mathrm{sign}\bigl(\mu[\phi]\bigr)\,\tilde\mu[\phi]} \; .
\end{equation}
In some cases this may work. However, the random walk guided by $|\mu[\phi]|$ is very likely
to predominantly sample unimportant regions in phase space.

At this time no completely satisfactory solution to the sigh problem
exists, although there are some promising attempts including positive
projection \cite{pp1,pp2}, fractal decomposition scheme \cite{fd}, and
Berry's phase \cite{bp} and its Stiefel manifold \cite{sm}. We believe
that the techniques discussed in the present work lay the grounds
for another possible solution path. The general success of the methods
studied here will depend on the development of more computationally
efficient algorithms than the ones used to illustrate the simple
examples treated below.
\paragraph{Smoothing out the Measure}
The idea behind the mollification technique is to use a convolution in order to smooth
out and filter the measure: the highly oscillatory measure given by $e^{i\,
  S[\phi]}\, \mathcal{D}\phi$ is convoluted with some suitable function (called
\emph{a mollifier}; see Appendix \ref{sec:background} for more details) and as a
result only an effective contribution is left (rapid oscillations of
the measure cancel out when integrated over the slowly varying mollifier).

A convolution is an integral that expresses the amount of overlap of one function $g$ as it
is shifted over another function $f$. It therefore ``blends'' one function with
another. In mathematics, mollifiers are smooth functions with special properties, used in
distribution theory (generalized functions) to create a sequence of smooth functions
approximating non-smooth functions, via a convolution.

In an  application of this idea to QFT we will replace the rapidly oscillating
measure in Minkowski space $(I = \exp\{i\, S[\phi]\})$ by smooth functions
$(I_{\epsilon} = \conv{\eta_{\epsilon}}{I})$ so that we can run numerical
simulations that are otherwise prohibitive (see \ref{sec:background}). At this
stage, the only requirement that is made is that: $\int \eta_{\epsilon}(x)\, dx
= 1$, where the integral is performed over the domain of $\eta_{\epsilon}$. The
parameter $\epsilon$ controls the approximation, the smoothness, being made, and
in the $\epsilon\rightarrow 0$ limit, the original expressions are
recovered. (The meaning of this $\epsilon$ parameter is that of dilating or
contracting the mollifier itself, thus controlling the range over which the
filtering is done. Note that the calculations should be done using a
non-vanishing value for $\epsilon$ in order to avoid the analytical expressions
which are hard to handle numerically.)
\section{Mollifying Quantum Field Theory} \label{sec:qft}
The technique described above (and detailed in \ref{sec:background}) can be
applied to [Lattice] QFT by introducing the mollified path integral

\begin{equation}
  \label{eq:mollqft}
  \mathcal{Z}_{\epsilon}[J] \equiv \varint \biggl\{\varint \eta_{\epsilon}[\phi - \varphi]\,
    e^{i\, S[\varphi;J]}\, \mathcal{D}\varphi\biggr\}\, \mathcal{D}\phi \; .
\end{equation}
This form is useful to set up numerical computations. Performing the
mollification before performing the path integral can be very advantageous.

Mollifying the integrand (i.e., the complex exponential of the action)
smooths the highly oscillatory integral. The integrand changes its form, from
the canonical $I$ to the mollified $I_{\varepsilon}$, in the following fashion,

\begin{equation*}
  I[\phi;J] = e^{i\, S[\phi;J]} \; \longmapsto\; I_{\epsilon}[\varphi; J] =
    (\conv{\eta_{\epsilon}}{I})[\varphi; J] = \varint
    \eta_{\epsilon}[\varphi - \phi]\, e^{i\, S[\phi;J]}\, \mathcal{D}\phi \; ,
\end{equation*}
i.e., the convolution with $\eta_{\epsilon}$ changes variables: $\phi \mapsto
\varphi$. Taking [functional] derivatives of $\mathcal{Z}$ (which
yield Green's functions) is just the same as taking them with respect to
$\mathcal{Z}_{\epsilon}$, because the derivative operator commutes with the
mollification.

The [Feynman] Path Integral is constructed just as before:

\begin{align}
  \nonumber
  \mathcal{Z}[J] &= \varint I[\phi;J]\, \mathcal{D}\phi \; ; \\
  \nonumber
  \mathcal{Z}_{\epsilon}[J] &= \varint I_{\epsilon}[\varphi;J]\, \mathcal{D}\varphi\; .\\
  \intertext{Note that,}
  \nonumber
  \mathcal{Z}_{\epsilon}[J] &= \varint\Biggl\{ \underbrace{\varint \eta_{\epsilon}[\varphi
    - \phi]\, e^{i\, S[\phi;J]}\, \mathcal{D}\phi}_{I_{\epsilon}[\varphi;J]} \Biggr\}\,
    \mathcal{D}\varphi\; . \\
    \label{eq:zeqze}
  \therefore\; \mathcal{Z}_{\epsilon}[J] &\equiv \mathcal{Z}[J] \; ;
\end{align}
where standard properties of convolutions (see \cite{dtta}) have been used in the last
step.

Even though the above result is analytic, when one goes to the simulations, a
small dependence on $\epsilon$ shows up. (See Section \ref{subsec:tm} for a more detailed
discussion on this matter.)
\subsection{Importance Sampling}\label{subsec:isspm}
The next step consists of choosing an
appropriate sampling function (for the Monte Carlo simulation):

\begin{equation*}
  \mathcal{Z}_{\epsilon}[J] \equiv \varint \biggl\{\varint \eta_{\epsilon}[\phi - \varphi]\,
    \frac{e^{i\, S[\varphi;J]}}{W[\varphi]}\, \mathcal{D}W\biggr\}\, \mathcal{D}\phi \; ,
\end{equation*}
where $\mathcal{D}W = W[\varphi]\, \mathcal{D}\varphi$, i.e., just a change of
variables in order to make the computations more robust: it is possible to use the
integrand profile in order to speed up the calculation.

An appropriate choice for the importance sampling function, $W$, is usually
given by $W_{\epsilon}[\phi] = \big|I_{\epsilon}[\phi]\big|$, \cite{stcqp}.

A little digression is in order: If all we wanted to do was to simulate a
certain [Euclidean] QFT on a lattice, the above reasoning would be just as
valid, modulo the mollification, i.e., we would conclude that the proper
importance sampling function was $W[\phi] = \big|I[\phi]\big|$. The full-fledged
formula is useless for doing computations since it implies full knowledge of
the theory being calculated. A better choice is a simple approximation: we
choose a saddle-point approximation in order to make the expression more
manageable and still keep most of the characteristics of the integrand.

Thus, the importance sampling function ($W[\phi]$), which is used just to better
guide the simulation, would be nothing but the absolute value of the sum over
all saddle-points of the theory in question. Assuming we have only one such
saddle-point, the well known answer is given by:

\begin{align}
  \nonumber
  W[\phi] &\approx \Biggl| \frac{e^{i\, S[\phi_0]}}{\hksqrt{\det\bigl\{-\partial^2 + m^2 +
    \mathcal{V}''[\phi_0]\bigr\}\,{}}} \Biggr| \; ; \\
  \nonumber
  &= \Bigl| e^{i\, S[\phi_0]}\Bigr|\, \Bigl| e^{-\frac{1}{2}\,
    \mathrm{tr}\left\{\log(-\partial^2 + m^2 + \mathcal{V}''[\phi_0])\right\}} \Bigr|\;;\\
  \label{eq:sadw}
  &= e^{-\frac{1}{2}\, \mathrm{tr}\{\log(-\partial^2 + m^2 + \mathcal{V}''[\phi_0])\}}
    \equiv \Bigl(\det\bigl\{-\partial^2 + m^2 + \mathcal{V}''[\phi_0]\bigr\}\Bigr)^{-1/2}\;.
\end{align}

Using this to find an importance sampling function for a simple scalar QFT
on the lattice requires solving for the determinant above. In Lattice QCD, the 
problem of the \emph{fermion determinant} is a very time-consuming
operation that must be done for \emph{every} step of the MC calculation.

But this is not the only computational bottleneck. The more severe one comes from the so
called \emph{sign problem}: The exponent in $I[\phi]$ is not bounded from below,
therefore one cannot guarantee the ergodicity of the Markov Chain underlying the Monte
Carlo draws, i.e., a much bigger number of draws would be needed in order to
yield any meaningful answer. This is the reason why calculations are done in
Euclidean space rather than Lorentzian/Minkowski space. In Euclidean space
(Wick-rotating the integrand), the exponent in question becomes $I_E[\phi] =
e^{-S[\phi]}$: it is bounded from below and the Monte Carlo method works fine
(the Markov Chain behind it becomes ergodic).

It is to tackle this problem that the method of mollifiers comes in: mollifying $I[\phi]$
will yield a smooth function, $I_{\epsilon}[\phi]$, whose properties help the
convergence of the MC computation. Furthermore, if this can really be accomplished and
calculations with imaginary exponents become a reality, the next logical step is to
analyze the different phases of the given QFT. As shown in \cite{tvbcsde}, the different
phases of a theory can be picked out with a mere choice of boundary conditions, which is
the same as properly defining the measure of the [Feynman] Path Integral. The
upshot is that if we want to obtain answers other than the ones that can
be reached via perturbation theory, either the measure of the path integral or
the boundary conditions of the Schwinger-Dyson equations have to be modified
\cite{tvbcsde}.

Once there is nothing preventing the measure of the [Feynman] Path Integral,
from being (in the most general case) complex-valued --- in order to account for
the phase structure of the theory ---, the fact that its integrand is also
complex-valued has to be taken more seriously, afterall this is a
non-perturbative result: a perturbative series only works if we know, \emph{a
  priori}, in which phase we are working, so it can be tailored to that
particular sector of the theory.

Thus, the need to address the sign problem becomes even clearer: it is not
simply a problem of doing MC in Lorentzian/Minkowski space, it is mostly a
problem of being able to compute all possible solutions --- all the different
phases --- of a given QFT.
\subsection{Mollifying the Importance Sampling Function}\label{subsubsec:mis}
As outlined before, the idea is to use the mollification technique to handle
the highly oscillatory terms present in the importance sampling function.

There are \emph{three} possible choices for the sampling function $W$:

\begin{enumerate}
\item Taylor-expand and mollify the complex integrand: \textit{no} knowledge about the
  saddle-points is necessary, but it is very computer intensive, which is the
  reason we will not focus on it;
\item As done in \eqref{eq:fieldw}, where the integrand is saddle-point expanded and
  mollified, but the \textit{explicit} form of the mollifier is used in order to perform
  the remaining integral: not so heavy on the computer, but some knowledge about the
  saddle-points is needed;
\item As done in \eqref{eq:mollw}, where the integrand is saddle-point expanded and
  mollified, generalizing the standard saddle-point approximation: less intensive of all
  3, but even more knowledge about the saddle-points is needed.
\end{enumerate}

The difference between the second and third methods above is that the integral
in \eqref{eq:2ndordergradint} is explicitly carried out in the former, but not in
the latter. As explained below, the third method (listed above) consists of the
simple generalization of the well known saddle-point approximation; however, we
need to note that more knowledge about the particular saddle-points is needed in
order to perform the contour integration involved in this scheme (and this is
the difference with respect to the second method).

The calculation below shows the second method outlined above, known as
\emph{2\,\raisebox{0.9ex}{\tiny nd}-order gradient approximation}:

\begin{align}
  \nonumber
  W_{\epsilon}[\phi] &= \big|I_{\epsilon}[\phi]\big| = \left| \varint
    \eta_{\epsilon}[\phi - \varphi]\, e^{i\, S[\varphi]} \,
    \mathcal{D}\varphi\right| \; . \\
  \intertext{Now, a saddle-point expansion is performed on the action,}
  \nonumber
  W_{\epsilon}[\phi] &\approx \left| \varint \eta_{\epsilon}[\phi - \varphi]\,
    e^{i\, \left(S[\varphi_0] + \frac{1}{2}\, (\varphi - \varphi_0)^2\,
        S''[\varphi_0]\right)} \mathcal{D}\varphi \right| \; . \\
  \intertext{A mollifier needs to be chosen, and for the purposes of this
  calculation, a \emph{Gaussian} one is our choice:}
  \nonumber
  \eta_{\epsilon}[\phi - \varphi] &= e^{-\frac{1}{2}\, (\phi -
    \varphi)^2/\epsilon^2} \; ;\\
  \label{eq:2ndordergradint}
  W_{\epsilon}[\phi] &= \left| \varint_{-\infty}^{\infty} e^{-\frac{1}{2}\,
    (\phi - \varphi)^2/\epsilon^2}\, e^{i\, \left(S[\varphi_0] + \frac{1}{2}\,
      (\varphi - \varphi_0)^2\, S''[\varphi_0]\right)} \mathcal{D}\varphi
      \right| \; ; \\
  \label{eq:fieldw}
  \Rightarrow\; W_{\epsilon}[\phi] &= \left| \hksqrt{2\, \pi\, \epsilon^2\xspace}\,
    \frac{\exp\left\{i\, S[\varphi_0] + \frac{i}{2}\, \frac{(\phi - \varphi_0)\,
    S''[\varphi_0]\, (\phi - \varphi_0)}{1 - i\, \epsilon\, S''[\varphi_0]\,
    \epsilon}\right\}}{\hksqrt{\det\big\{1 - i\, \epsilon\, S''[\varphi_0]\,
    \epsilon\big\}\xspace\,}}  \right| \; .
\end{align}

The important point to note about the derivation above is that the \emph{explicit}
form of the mollifier, $\eta_{\epsilon}$, had to be used and, because of 
that and the nature of the MC simulation, we do not necessarily need a good knowledge
about $\varphi_0$. This means that we can compute the above importance sampling function,
\eqref{eq:fieldw}, and use a \emph{trial} $\varphi_0^{\text{trial}}$: the [MC] simulation
will do the job of moving towards the exact $\varphi_0$ and pick out the different
phases of the theory. As shown elsewhere, \cite{nspmc,rtnnm,mmc,stcqp}, this works quite
fine.

However, for more relevant cases (e.g. 4-dimensional QFTs) the time required for
simulations becomes unrealistic. The bottom-line is that a simple-minded
implementation of the method will have the code/com\-put\-er doing the work of
mapping the $(\phi,\varphi)$-space, while one could use that information
beforehand in order to speed things.

The way out of this is to generalize the following well known result:
$\mathcal{I}(s) = \int_{\mathds{C}} g(z)\, e^{s\, f(z)}\, dz = \hksqrt{2\,
  \pi\,{}}\, g(z_0)\, e^{s\, f(z_0)}\, e^{i\, \alpha}/\hksqrt{|s\, f''(z_0)|}\,
;\; s\in\mathbb{R}, \; z\in\mathbb{C}$; i.e.,

\begin{equation}
  \label{eq:mollw}
  W_{\epsilon}[\phi] = \left| \hksqrt{2\, \pi\,{}}\, \frac{\eta_{\epsilon}[\phi - \varphi_0]\,
    e^{i\, S[\varphi_0]}}{\hksqrt{\left|\det\bigl\{S''[\varphi_0]\bigr\}\right|\,{}}}\right| \; .
\end{equation}

At this point, the similarity between \eqref{eq:sadw} and
\eqref{eq:mollw} is clear, in fact, they are the same formula, except that, in
the former case there is no mollification while in the latter there is
mollification.
\section{Tuning the Mollification} \label{subsec:tm}
At this point we are almost done with the analysis of the mollification technique, the
only aspect remaining being the optimal choice of $\epsilon$, the parameter that regulates
the approximation.

From the discussion in appendix \ref{sec:background}, it is clear that in the
$\epsilon \rightarrow 0$ limit we recover the original theory. However,
numerically, this is not exactly what happens, because of numerical and
statistical fluctuations. Therefore, the choice of an optimal value for
$\epsilon$ is crucial.

Using an information theoretic viewpoint, we can choose $\epsilon$ to
optimally compress $W_{\epsilon}[\phi]$ around each stationary phase
point. Therefore, we can define an information entropy based on
$W_{\epsilon}[\phi]$ in the following way:

\begin{equation}
  \label{eq:infentropy}
  \mathcal{S}[\epsilon] = - \frac{ \varint W_{\epsilon}[\phi]\, \log\bigl(
    W_{\epsilon}[\phi]\bigr) \, [d\phi]}{\bigl(\varint W_{\epsilon}[\phi]\, [d\phi]\bigr) +
    \log\bigl(\varint W_{\epsilon}[\phi] \, [d\phi]\bigr)} \; .
\end{equation}
The optimal $\epsilon$ is the one that minimizes $\mathcal{S}[\epsilon]$. This
value corresponds to maximal compression of the information in the importance
sampling function.

In order to illustrate the above, let us consider a very simple [0-dimensional] example,
given by $S[\phi] = m^2\, \phi^2/2$. In this case, the saddle-point is $\phi_0=0$ and the
importance sampling function and the information theoretic entropy are given by:

\begin{align*}
  W_{\epsilon}[\phi] &= \frac{e^{-\phi^2/2\epsilon^2}}{\hksqrt{2\,\pi\,\epsilon^2}}\,
    \frac{e^{i\, S[\phi_0]}}{\hksqrt{S''[\phi_0]}} \; ;\\
  &= \frac{e^{-\phi^2/2\epsilon^2}}{m\, \epsilon\, \hksqrt{2\, \pi}} \; ;\\
  \mathcal{S}[\epsilon] &= \frac{\log\bigl(m\, \epsilon\,
    \hksqrt{2\, \pi}\bigr) - \epsilon^2}{1 - m\, \log(m)} \; .
\end{align*}

Plotted below are the graphs of the importance sampling function (also showing how it
varies with $\epsilon$) and of the information theoretic entropy (both using $m = 1$):
\begin{center}
  \hfill
  \includegraphics[scale=0.7]{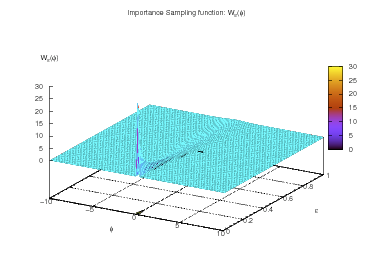}
  \hfill
  \includegraphics[scale=0.47]{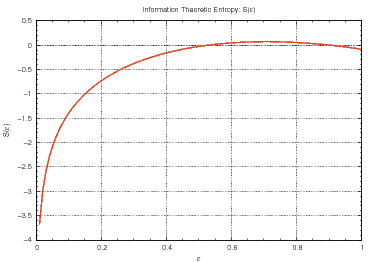}
 
  {\footnotesize \noindent\textbf{Figure 1:} Importance Sampling function
    (leftmost plot, $m = 1$, $W_{\epsilon}[\phi] = e^{-\phi^2/2\epsilon^2}/\epsilon\,
    \hksqrt{2\, \pi}$) and information theoretic entropy (rightmost plot, $m =
  1$, $\mathcal{S}[\epsilon] = \log\bigl(\epsilon\, \hksqrt{2\, \pi}\bigr) -
  \epsilon^2$) for the action $S[\phi] = m^2\, \phi^2/2$.}
\end{center}

We can clearly see the importance function peaking around $\phi=0$, which is the expected
behavior (\textit{i.e.}, it happens around the saddle-point $\phi_0=0$), while the
information theoretic entropy shows that the best parameter for the
mollification is $\epsilon \rightarrow 0$: this is not desirable, once
$\lim_{\epsilon \rightarrow 0}\mathcal{S}(\epsilon) \rightarrow -\infty$, but it 
simply shows that we should use a small non-vanishing value for $\epsilon$, as
mentioned before in Section \ref{sec:intro}.
\section{Simulation Details}\label{subsec:std}
In order to perform the simulation as outlined in \ref{subsubsec:mmls}, some technical
details had to be taken into more serious account. Among them, the important ones are:
random-number generation and non-local algorithms. Those will be the bottle-necks, as
noted in \cite{selcrngws}.
\paragraph{Random Number Generators}
Even though some consider the random-number question a solved one, this is not
always the case \cite{selcrngws}. For this reason, the choice of which
random-number generator (RNG) to use is still a critical one, not only for speed reasons
but because of systematic errors as well. In order to address this issue the choice made
was for the Mersenne Twister RNG \cite{mtrng} (period of $2^{19937} - 1$).
\paragraph{Non-local Algorithms}
As for non-local algorithms, the choice made was for a Genetic Algorithm (GA)
\cite{ugosumcmci}. The ``population'', in the Lattice QFT sense, consists of the lattice
(\textit{i.e.}\/ spacetime) points. For a given ``recombination rate'', an initial
population gets \emph{randomly} divided into pairs of [lattice] points and
``crossing-over'' (or ``genetic'') operators are applied to these pairs in order to
generate different sets of pairs. In this context, the crossing-over operators exchange
the coordinates of given points, thus there are 4 of those operators, namely:
$T_t,\, T_x,\, T_y\,, T_z$. That is, if the crossing-over operator $T_x$ is
applied to the points $\phi = (\phi_t, \phi_x, \phi_y, \phi_z)$ and $\phi' =
(\phi'_t, \phi'_x, \phi'_y, \phi'_z)$ the outcome will be: $T_x [\phi,\phi'] =
\{(\phi_t, \phi'_x, \phi_y, \phi_z),\, (\phi'_t, \phi_x, \phi'_y,
\phi'_z)\}$. Analogous definitions are valid for the other crossing-over
operators. The key features for this choice were: which $T_i$ to use is random;
whether or not the change $\phi \mapsto \phi'$ is made depends on a given
probability and the recombination rate can be arbitrarily chosen (although
keeping it below 1\% showed to be a good tune). Moreover, the genetic operators
are unitary, i.e., $T_i = T_i^{-1}$. This implies that $T_i^2 = \mathds{1}$,
which guarantees the so-called \emph{detailed balance} of the MC simulation
(this is the \emph{ergodicity} of the algorithm). At this point, the only step
remaining is the explanation of the \emph{probability profile} used in algorithm
\ref{alg:gl} below.

\begin{algorithm}
  \caption{Genetic Algorithm}
  \label{alg:ga}
  \begin{algorithmic}[1]
    \STATE Choose the recombination rate such that: $0 \leqslant r\leqslant 1$ \COMMENT{usually
      between $0.5\%$ and $1.0\%$}
    \STATE Draw a random number $\beta\in [0,1]$ and compare with $r$
    \IF{$\beta \geqslant r$}
      \STATE Metropolis Monte Carlo loop \COMMENT{using the \emph{random walk} technique}
    \ELSE
      \STATE Genetic loop \COMMENT{see algorithm \ref{alg:gl} below}
    \ENDIF
  \end{algorithmic}
\end{algorithm}
\begin{algorithm}
  \caption{Genetic Loop}
  \label{alg:gl}
  \begin{algorithmic}[1]
    \STATE Choose a random pair: $\{\phi_1, \phi_2\} = \{\phi(t_1,x_1,y_1,z_1),\,
      \phi(t_2,x_2,y_2,z_2)\}$
    \STATE Generate the pair $\{\phi'_1, \phi'_2\} = \{\phi(t'_1,x'_1,y'_1,z'_1),\,
      \phi(t'_2,x'_2,y'_2,z'_2)\} = T_i(\phi_1, \phi_2)$
    \STATE Draw a random number $c\in [0,1]$
    \IF{$c > P(\phi'_1, \phi'_2)/P(\phi_1, \phi_2)$}
      \STATE do \textbf{nothing}
    \ELSE
      \STATE perform the exchange $(\phi_1, \phi_2) \mapsto (\phi'_1, \phi'_2)$
    \ENDIF
  \end{algorithmic}
\end{algorithm}

In the notation used below, $P$ is such probability profile
and $P(\phi'_i,\phi'_j)$ means that this profile is calculated using the [GA generated]
points $\phi'_i$ and $\phi'_j$. It proved useful to chose the mollified
importance sampling function as this profile $P$. Also, note that $\phi_i$ and
$\phi_j$, just like $\phi'_i$ and $\phi'_j$, are just particular values of the
field $\phi$ at the sites $i$ and $j$. A comparison between these two profiles
is then performed.
\subsection{Euclidean Lattice Simulations}\label{subsubsec:els}
In order to be able to understand better what will happen in the
Lorentzian/Minkowski case, a quick and dirty reminder of the Euclidean one is
presented below.

In what follows, $\Phi$ is the set of all possible field configurations and $||\Phi||$ is its
\emph{cardinality} (i.e., the number of elements in the set). For each element of $\Phi$
(i.e., for each field configuration) one has to average the observable over all
the lattice points (which are computed via Metropolis). Each element in the Markov Chain
(generated by the Metropolis MC algorithm) is denoted by $\phi^{[i]}$. To obtain the final
result, an average over all field configurations is made. (To make clear that the
final goal is the actual lattice computation, the measure is denoted as $[d\phi]$.)

\begin{align*}
  \langle O \rangle &= \frac{\varint O[\phi]\, \exp\bigl\{-S[\phi]\bigr\} \,
    [d\phi]}{\varint \exp\bigl\{-S[\phi]\bigr\} \, [d\phi]} \; , \\
  &= \frac{1}{||\Phi||}\, \sum_{\phi\in\Phi}
    \Biggl\{\frac{\Bigl[\frac{\text{Volume}}{N^d}\Bigr]\cdot \sum_{i=1}^{N^d}
    O[\phi^{[i]}]}{\Bigl[\frac{\text{Volume}}{N^d}\Bigr]\cdot N^d}\Biggr\} \; , \\
  \therefore\; \langle O \rangle &= \frac{1}{||\Phi||}\, \sum_{\phi\in\Phi} \Biggl\{
  \frac{1}{N^d}\, \sum_{i=1}^{N^d} O[\phi^{[i]}]\Biggr\} \; .
\end{align*}
\subsection{Mollified Minkowski Lattice Simulations}\label{subsubsec:mmls}
In the Minkowski version of the above, the Wick rotation is not performed, therefore we
are left with the original form of the exponent,

\begin{equation*}
   \langle O \rangle = \frac{\varint O[\phi]\, \exp\bigl\{i\, S[\phi]\bigr\}
    \, [d\phi]}{\varint \exp\bigl\{i\, S[\phi]\bigr\} \, [d\phi]} \; .
\end{equation*}

The above functional will be mollified in order to yield more tractable
expressions. The fact that there will be two functional integrations (rather than just
one, like above) should not bring many problems. The real question here stems from the
fact that the mollifier, $\eta$, \emph{mixes} the two of them.

\begin{align}
  \nonumber
  \langle O \rangle &= \frac{\varint\Bigl\{\varint \eta_{\epsilon}[\phi
    - \varphi]\, O[\varphi]\, \exp\bigl\{i\, S[\varphi]\bigr\} \, [d\varphi]\Bigr\}
    [d\phi]}{\varint\Bigl\{\varint \eta_{\epsilon}[\phi - \varphi]\,
    \exp\bigl\{i\, S[\varphi]\bigr\} \, [d\varphi]\Bigr\} [d\phi]} \; ; \\
  \label{eq:mollobs}
  \therefore\; \langle O \rangle &= \frac{1}{||\Phi||}\, \frac{1}{||F||}\,
  \sum_{\substack{\phi\in\Phi \\ \varphi\in F}}\, \Biggl\{\frac{\sum_{i,j=1}^{N^d}\;
    \eta_{\epsilon}[\phi^{[j]} - \varphi^{[i]}]\, O[\varphi^{[i]}]\, \exp\bigl\{i\,
    S[\varphi^{[i]}]\bigr\} / W_{\epsilon}^{\Phi}[\varphi^{[i]}]}{\sum_{i,j=1}^{N^d}\; \eta_{\epsilon}[\phi^{[j]}
    - \varphi^{[i]}]\, \exp\bigl\{i\, S[\varphi^{[i]}]\bigr\} /
    W_{\epsilon}^{\Phi}[\varphi^{[i]}]}\Biggr\} \; ;
\end{align}
where it is understood that the configurations in $\Phi$ are chosen with respect
to the importance sampling function, $W_{\epsilon}^{\Phi}[\phi]$, and the
configurations in $F$ are chosen with respect to a \emph{uniform}
distribution. This happens because the mollification process (namely the
\emph{convolution}) mixes the variables from $\Phi$ and $F$ together. Therefore,
the only way to implement this ``mixing'' is by having a uniform distribution
for $F$ and implementing the ``interaction'' with the variables in $\Phi$ ---
via the mollification --- explicitly, using $\eta$. Note that $||\Phi|| = ||F||$.
\section{Spontaneous Symmetry Breaking}\label{subsec:symb}
Note that in this paper we use the designation ``spontaneous symmetry breaking'' somewhat
loosely in that we apply this term to all solutions which do not approach those of coupling constant
perturbation theory in the limit where the coupling vanishes.
We have been discussing lattice QFT in Lorentzian/Minkowski spacetime because one
of the main objectives of this paper is to examine a numerical
approach to directly calculate Green's functions in different phases
in QFT. This necessitates being able to evaluate path integrals of
complex exponentials. The arguments for this have been given elsewhere \cite{tvbcsde}
and will be illustrated in a particular example in the following section,
but it is fairly easy to understand why this is the case. The
traditional [Feynman] Path Integral formulation of a QFT involves integration of
the exponential of the action over every field variable at every space
time point. Usually, these integrations (assume the action is
written in terms of self adjoint fields) range from negative to
positive infinity along the real axis. In the limit of small
couplings, this form of the [Feynman] Path Integral generates perturbation
theory and the results appear (for finite number of expansion terms) to
be regular at vanishing coupling. Thus, all expansions that are not
regular at vanishing coupling, such as the traditional symmetry
breaking expansion of quartic scalar field couplings, are excluded. In
order to avoid this restriction and produce all possible solutions of
the QFT it is necessary to extend the [Feynman] Path Integral integrations to
complex values of the fields in a way consistent with the field
equations and reality properties of the theory.

Indeed, as shown in \cite{tvbcsde} and section \ref{subsec:r}, the
different phases of the theory emerge through the varied boundary
conditions consistent with the equations of motion. Equivalently,
rather than varying the boundary conditions of equations of motion,
the measure of the Path Integral can be changed. In general the number
of choice of paths of integration for the path integral correspond to
the number of independent solutions of the Schwinger-Dyson differential equations.

If we define a QFT via its [Feynman] Path Integral, all we need to know is the action,
$S[\phi, J]$ (for a field $\phi$ whose source is $J$), of a given model, for then we can
write the generating functional as:

\begin{equation*}
  \mathcal{Z}[J] = \mathcal{N}\, \varint\exp\bigl\{i\, S[\phi; J]\bigr\}\,
    \mathcal{D}\phi\; ,
\end{equation*}
where $\mathcal{N}$ is a normalization constant such that $\mathcal{Z}[J=0] =
1$. The crucial question that remains unanswered in this approach is: \emph{``How does one
  properly define the measure $\mathcal{D}\phi$?''}

The best answer so far (for 4-dimensional systems) says that this can only be
done for \textit{free} QFTs, via the use of cylindrical functions
\cite{Baez,Ashtekar}. (Note that this only happens in the
continuum. In its Lattice formulation, QFT is free from such peculiarities
because of the lattice regularization.)

Analogously, defining a QFT via its Schwinger-Dyson equation,

\begin{equation*}
  \frac{\delta S[-i\, \tfrac{\delta}{\delta J}]}{\delta\phi}\mathcal{Z}[J] - J(x)\,
    \mathcal{Z}[J] = 0 \; ;
\end{equation*}
is equivalent to substituting $\phi \mapsto -i\, \tfrac{\delta}{\delta J}$ in the
action, requiring the equations of motion to be the solutions that extremize it. Note
that the Schwinger-Dyson equations are a system of infinitely many [partial] differential
equations, one per each point of spacetime.

Therefore, when thinking in terms of differential equations, the boundary conditions are
responsible for the phase structure of the theory. Intuitively, the picture that comes to
mind is that of a portion of space divided into as many subsets as there are solutions to
our equations of motion, such that in each of those regions, the equations of motion
satisfy appropriate boundary conditions.

On the other hand, when thinking about the [Feynman] Path Integral, usually it does not
seem bothersome that the measure is not properly well-defined. In fact, determining the
boundary conditions for the Schwinger-Dyson equations is analogous to determining the
measure for the Path Integral. Thus, just like the boundary conditions, the measure is
responsible for the phase structure of the theory (in the integral representation of the
problem).

For completeness sakes, this is how a QFT in different spacetime dimensions, $d$, behaves:

\begin{description}
 \item[$\boldsymbol{d = 0}$] For 0-dimensional QFTs --- i.e., QFT on a point (the
   [Feynman] Path Integral degenerates into a simple integral) --- there is no
   such thing as phase transition since there is no such thing as
   dynamics. However, the phase structure of the theory survives, given by the
   different boundary conditions (resp. measure) needed in order to determine all
   the solutions to the equations of motion.
 \item[$\boldsymbol{d = 1}$] For 1-dimensional QFTs --- i.e., Quantum
   Mechanics ---, again, there is no such thing as phase transitions, for if the
   theory has 2 different vacua we could take a linear combination of them to be
   the ``real'' vacuum state, given that they would be related by tunneling. (For
   the analogous case in Condensed Matter physics, please refer to \cite{llsp1}.)
 \item[$\boldsymbol{d \geqslant 2}$] In this case, phase structure and transition
   exist; this is the complete scenario. (Note that there is no operator that
   relates 2 inequivalent $\theta$-vacua of the theory, therefore they belong to
   different algebras.)
\end{description}

The real lesson to be learned from all of this is that Spontaneous Symmetry Breaking is a
phenomena generated by boundary conditions; whether you use them to define the measure of
the Path Integral or to define the Schwinger-Dyson equation is just a matter of personal
preference. It is not said anywhere that the limits of the Path Integral have to be real;
what we need to have are real observables. In fact, as shown below on section
\ref{subsec:r}, it turns out that in order to have symmetry breaking we need a
measure that is not necessarily real: this will enable the computation of all the
solutions of a given QFT \cite{tvbcsde}.
\section{Results}\label{subsec:r}
In what follows, the results obtained thus far are presented. In short, they are in 0,
1 (time) and 4 spacetime dimensions.
\subsection{Lower Dimensional}\label{subsubsec:ld}
Let us start by addressing the 0- and 1-dimensional results. Plainly and simply put, this
means that we are solving a simple integral in 0 spacetime dimensions since the
[Feynman] Path Integral degenerates into a standard integral and, in the 1-dimensional
(time) case, one will be doing Quantum Mechanics.

For the 1-dimensional results, refer to \cite{stcqp}. There, the Quantum
Chemistry of the problem is fully treated and addressed. Note, however, that the
importance sampling function chosen in \cite{stcqp} is different than the one
used in this work: the quantum chemistry was done using \eqref{eq:fieldw} while we use
\eqref{eq:mollw}.

Below, the 0-dimensional results are summarized and, in order to illustrate the features
of the mollifier technique, different properties of those models are made explicit.
\paragraph{Airy Function}
The action for this model is given by: $S(x) = x^3/3 + J\, x$. This is an interesting
model because we can explicitly calculate the partition function and compare it with the
results coming from the Mollified Monte Carlo procedure. Moreover, this model has 2
stationary phase points and the results displayed are from the one in the complex plane
which is not accessible with normal Monte Carlo. The quantity of interest is:
\hspace*{2em}
\begin{minipage}[t]{0.4\linewidth}
  \includegraphics[scale=0.3]{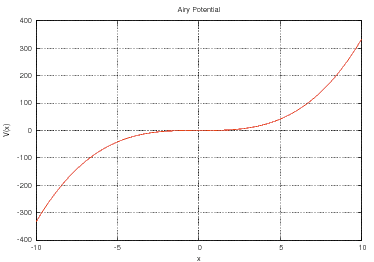}
\end{minipage}
\hspace{1em}
\begin{minipage}[b]{0.5\linewidth}
  {\footnotesize\noindent\textbf{Figure 2:} Plot of the Airy potential, $V(x) = x^3/3$,
    highlighting the fact that it is not bounded from below.}
\end{minipage}

\begin{equation*}
  \mathcal{Z}[J] = \frac{\int_{-\infty}^{\infty}\, \exp\Bigl\{i\,
    \frac{x^3}{3} + i\, J\, x\Bigr\}\, dx}{\int_{-\infty}^{\infty}\,
    \exp\Bigl\{i\, \frac{x^3}{3}\Bigr\}\, dx} \equiv \frac{\Ai(J)}{\Ai(0)} \; .  
\end{equation*}

The [first two] graphs below show the highly oscillatory behavior of the
integrand in the partition function: the left one is its the real part, while the
right one is its the imaginary part.

\begin{center}
  \hspace{\fill}
  \includegraphics[scale=0.55]{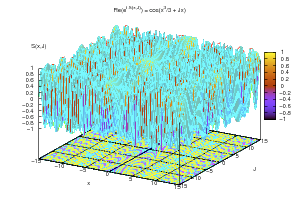}
  \hspace{\fill}
  \includegraphics[scale=0.55]{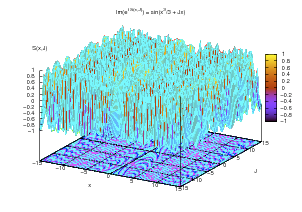}
  \hspace{\fill}

  \bigskip

  {\footnotesize\noindent\textbf{Figure 3:} Real and Imaginary parts of $e^{i\,
      x^3/3 + i\, J\, x}$ showing the highly oscillatory nature of the problem.}
\end{center}

Below we see the graphs of a particular 2-dimensional slice of the above pair,
where $J = -1$: the real (left) and imaginary (right) part of the mollified
Airy-integrand (in red, $\epsilon=0.1$) is in contrast to the non-mollified
integrand (green).

\begin{center}
  \includegraphics[scale=0.4]{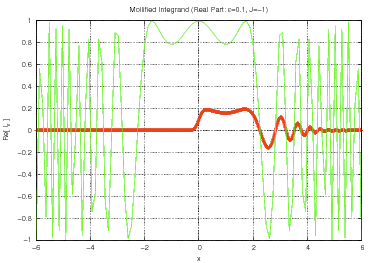}
  \hspace{2cm}
  \includegraphics[scale=0.4]{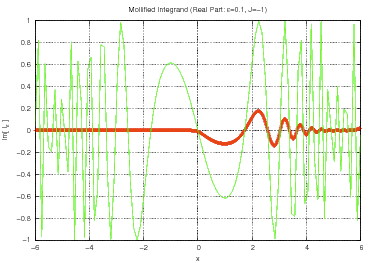}

  \bigskip \bigskip

  {\footnotesize\noindent\textbf{Figure 4:} Superimposed plots of mollified
    (red) and non-mollified (green) real and imaginary parts of the integrand
    $e^{i\, x³/3 - i\, x}$, showing the smoothness achieved on the $J=-1$
    two-dimensional slice of the previous graphs.}
\end{center}

Below we see the graphs of the real (left) and imaginary (right) part of the mollified
Airy-integrand for different values of the field $x$ and of the mollification
parameter $\epsilon$: analogous to Figure 3 but for the mollified version of the
integrand (which corresponds to the red lined plots of Figure 4).

\begin{center}
  \includegraphics[scale=0.5]{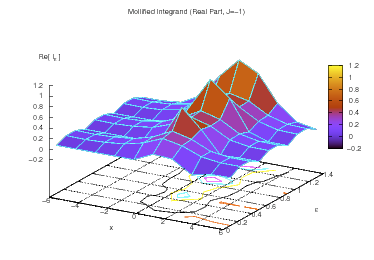}
  \hspace{\fill}
  \includegraphics[scale=0.5]{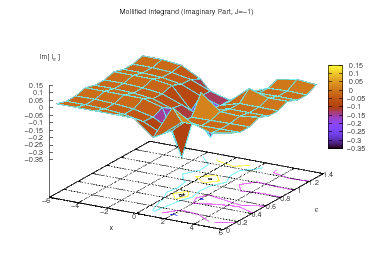}

  {\footnotesize\noindent\textbf{Figure 5:} Plots of the real and imaginary
    parts of the mollified integrand for various values of $x$ and $\epsilon$.}
\end{center}

The 2 stationary phase points are given by $x_0 \in
\bigl\{\pm\hksqrt{-J}\bigr\}$, where $x_0$ is the solution of $S'(x_0) = 0$, and the
[properly mollified] importance sampling functions are:

\begin{align}
  \label{eq:mollisairypos}
  W_{\epsilon}^{J \geqslant 0}[x] &= \exp\biggl\{-\frac{1}{2}\, \frac{x^2 - J}{\epsilon^2} \biggr\} +
    \exp\biggl\{-\frac{1}{2}\, \frac{x^2 + J}{\epsilon^2} \biggr\}
    \; ;\quad J \geqslant 0 \; ; \\
  \label{eq:mollisairyneg}
  W_{\epsilon}^{J < 0}[x] &= \exp\Biggl\{-\frac{1}{2}\, \biggl(\frac{x -
    \hksqrt{-J\,{}}}{\epsilon}\biggl)^2 \Biggr\} + \exp\Biggl\{-\frac{1}{2}\, \biggl(\frac{x +
    \hksqrt{-J\,{}}}{\epsilon}\biggl)^2 \Biggr\}\; ; \quad J < 0 \; ;
\end{align}
It is customary to see the source term above
called $t$ and identified with time but this is not the approach taken here.

Below we see the graphs of $W_{\epsilon}$ with respect to the $\epsilon$
parameter and $x$, for fixed values of the source: on the left $J=4$, and on the right
$J=-16$,

\begin{center}
  \hspace{\fill}
  \includegraphics[scale=0.55]{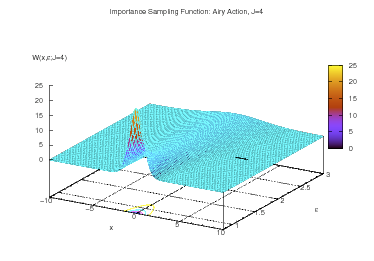}
  \hspace{\fill}
  \includegraphics[scale=0.55]{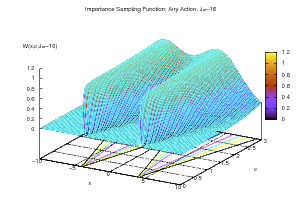}
  \hspace{\fill}

  {\footnotesize\noindent\textbf{Figure 6:} Plots of \eqref{eq:mollisairypos},
    for $J=4$, and \eqref{eq:mollisairyneg}, for $J=-16$.}

  \includegraphics[scale=0.45]{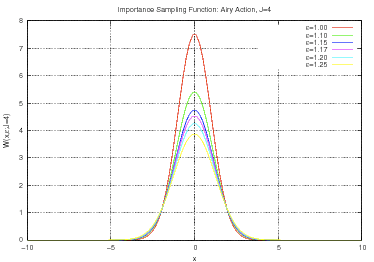}
  \hspace{\fill}
  \includegraphics[scale=0.45]{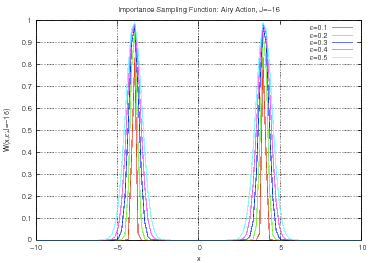}
  \hspace{\fill}

  \bigskip \bigskip

  {\footnotesize\noindent\textbf{Figure 7:} Plots of 2-dimensional slices of
    \eqref{eq:mollisairypos} and \eqref{eq:mollisairyneg} for various values of
    $\epsilon$.}
\end{center}

It is not difficult to see that there is an optimal value for the parameter
$\epsilon$, as shown below in the graph for the information theoretic entropy (see
Appendix \ref{sec:entropy} for further details): we have the information theoretic entropy
for $J\geqslant 0$ on the left, and $J<0$ on the right,

\begin{center}
  \includegraphics[scale=0.55]{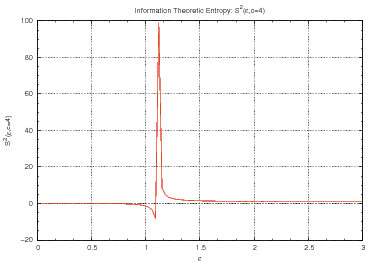}
  \hspace{\fill}
  \includegraphics[scale=0.55]{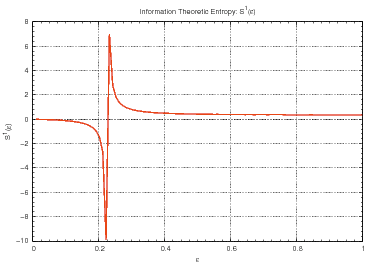}

  \bigskip \bigskip

  {\footnotesize\noindent\textbf{Figure 8:} Information theoretical entropy for
    finding the optimal value of the mollification parameter: on the left we
    have $J \geqslant 0$, and on the right we have $J < 0$.}
\end{center}

Furthermore, it is easy to perform the analytical calculations outlined above (using the
Airy functions on the Generating Functional, rather than its mollification) to get the
pure evaluation of this model. The results show the mollified answers (given a
proper choice of $\epsilon$) are sensationally accurate and show, already in this
simple example, that a region not allowable in ordinary Monte Carlo approaches can be
computed. In fact, due to the simplicity of this example, modern computers can
calculate it straightforwardly, without the need of special tricks to handle the highly
oscillatory integrand. The point of this example is twofold: show the precision and the
accuracy that the mollifier method can achieve and also to illustrate how this smoothing
procedure works.
\paragraph{0-dimensional $\boldsymbol\phi^4$ Theory}
Also known as ultra-local $\phi^4$, its action is given by: $S[\phi] = \mu\, \phi^2/2 +
g\, \phi^4/4$. As before, the saddle-point $\phi_0$ is such that $S'[\phi_0] = 0$. The
graphs for the Action above (positive mass on the left, negative mass on the right) are
given by:

\begin{center}
  \hfill
  \includegraphics[scale=0.5]{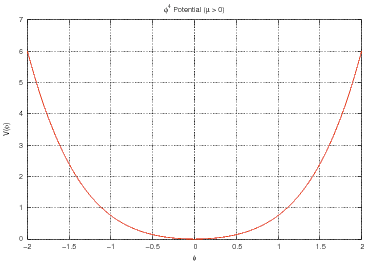}
  \hfill
  \includegraphics[scale=0.5]{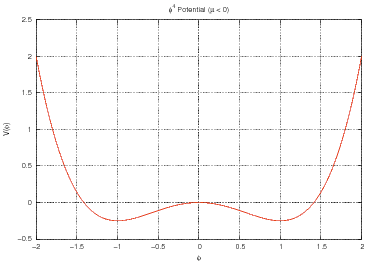}

  \bigskip \bigskip

  {\footnotesize\noindent\textbf{Figure 9:} $\phi^4$ potential for positive and
    negative values of $\mu$.}
\end{center}

Below, we show the highly oscillatory behavior of the integrand of the Partition Function
given by the above action, i.e., $I = e^{i\, (\mu\, \phi^2/2 + g\, \phi^4/4)}$, for
positive and negative values of the $\mu$ parameter:

\begin{center}
  \hspace{\fill}
  \includegraphics[scale=0.5]{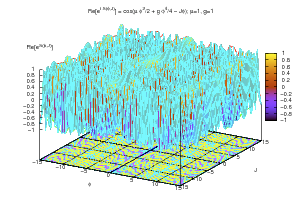}
  \hspace{\fill}
  \includegraphics[scale=0.5]{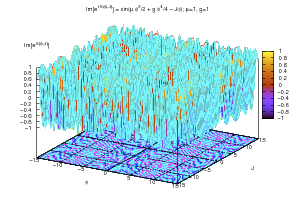}
  \hfill

  \hfill
  \includegraphics[scale=0.5]{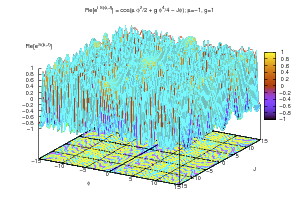}
  \hspace{\fill}
  \includegraphics[scale=0.5]{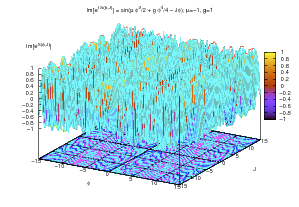}
  \hspace{\fill}

  \bigskip \bigskip

  {\footnotesize\noindent\textbf{Figure 10:} Plots of the Real and Imaginary
    parts of the integrand for various values of $J$ and $\mu = 1,\, g = 1$ and
    $\mu = -1,\, g = 1$.}
\end{center}

Below, we see the real (left) and imaginary (right) parts of the mollified
$\phi^4$-integrand (red) in comparison with their non-mollified counterparts (green):

\begin{center}
  \hfill
  \includegraphics[scale=0.5]{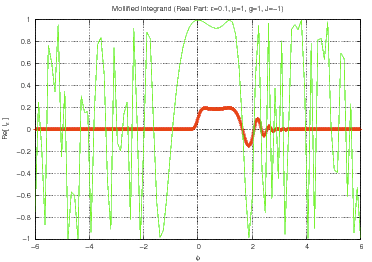}
  \hfill
  \includegraphics[scale=0.5]{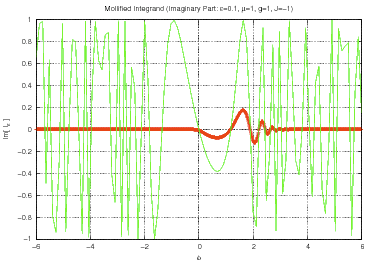}
  \hfill

  \bigskip \bigskip

  {\footnotesize\noindent\textbf{Figure 11:} Superimposed mollified (red) and
    non-mollified (green) real and imaginary parts of the integrand for $\mu
    =1,\, g = 1,\, J = -1,\, \epsilon = 0.1$.}
\end{center}

Below, we see the real (left) and imaginary (right) parts of the mollified
$\phi^4$-integrand for different values of the field $\phi$ and of the mollification
parameter $\epsilon$:

\begin{center}
  \hfill
  \includegraphics[scale=0.5]{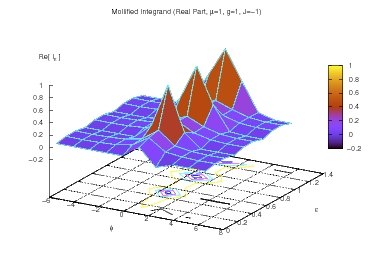}
  \hfill
  \includegraphics[scale=0.5]{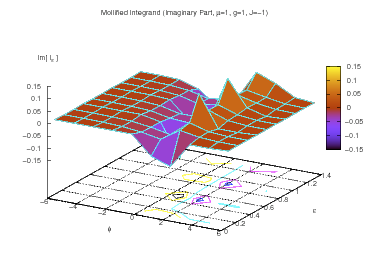}
  \hfill

  {\footnotesize\noindent\textbf{Figure 12:} Plots of the real and imaginary
    parts of the mollified integrand for various values of the field and the
    parameter $\epsilon$.}
\end{center}

As explained in \cite{tvbcsde,rtnnm}, the three different solutions/phases of this model
can be selected via an appropriate choice of boundary conditions: $\Gamma^0 = \mathbb{R}$,
$\Gamma^{+} = (-\infty,0) \cup (0, i\, \infty)$ and $\Gamma^{-} = (-\infty,0) \cup (0,
-i\, \infty)$. The leftmost figure below shows those boundaries in the Argand Plane:
$\Gamma^0 = \mathbb{R}$ is the green line, $\Gamma^{+} = (-\infty,0) \cup (0, i\, \infty)$
is the red line while $\Gamma^{-} = (-\infty,0) \cup (0, -i\, \infty)$ is the blue
line. The rightmost figure below shows the regions (shaded) of the complex $\phi$-plane,
$(\phi = \rho\, e^{i\, \theta})$, defined by $\cos(4\, \theta) \geqslant 0$. Any contour,
starting and ending at infinity, within one of these four domains corresponds to a
particular solution of the 0-dimensional $\phi^4$.

\begin{center}
  \begin{minipage}[b]{0.4\linewidth}
    \includegraphics{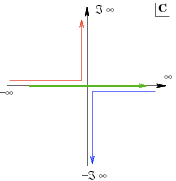}
  \end{minipage}
  \hspace{0.1\linewidth}
  \begin{minipage}[t]{0.4\linewidth}
    \includegraphics{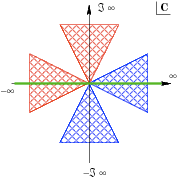}
  \end{minipage}
  \bigskip \bigskip

  {\footnotesize\noindent\textbf{Figure 13:} \textsc{Left:} Three different
    contours that render the Path Integral finite and depict the three distinct
    solutions obtained in this model. \textsc{Right:} Any contour that starts
    and finishes inside the same-color shaded regions (following the general
    directions shown on the left) corresponds to a particular finite solution.}
\end{center}

Using the three different contours above we are able to find the three different solutions
to the equation of motion of this 0-dimensional QFT. The Schwinger-Dyson equation is given by,

\begin{equation*}
  -i\, \biggl(\mu\, \frac{\delta}{\delta J} + g\, \frac{\delta^3}{\delta J^3}\biggr)\,
     \mathcal{Z}[J] = J \, \mathcal{Z}[J] \; ,
\end{equation*}
where we can easily see (again) that there has to be three solutions. These are
called \emph{Parabolic Cylinder Functions} and are denoted $U(\mu/g,J),\,
V(\mu/g,J),\, W(\mu/g,J)$. Those behave in the following manner (see Appendix
\ref{sec:pcf}):

\begin{description}
\item[Regular at $\boldsymbol{g \rightarrow 0}$] Consistent with perturbation theory,
\item[Singular $\boldsymbol{(\simeq\hksqrt{g})$ at $g \rightarrow 0}$] Symmetry Breaking,
\item[Singular $\boldsymbol{(\simeq\exp\{\mu/4\, g\})$ at $g \rightarrow 0}$] Instanton.
\end{description}

Thus,
\begin{equation*}
  \langle\mathcal{O}\rangle_{\epsilon}[J] = \frac{\varint_{-\infty}^{\infty}\,
    \varint_{\Gamma^{0,\pm}}\, \eta_{\epsilon}(\phi - \varphi)\, \mathcal{O}[\varphi]\,
    \exp\Bigl\{i\, \frac{\mu}{2}\, \varphi^2 + i\, \frac{g}{4}\, \varphi^4 - i\, J\,
    \varphi\Bigr\}\, d\varphi\, d\phi}{\varint_{-\infty}^{\infty}\,
    \varint_{\Gamma^{0,\pm}}\, \eta_{\epsilon}(\phi - \varphi)\, \exp\Bigl\{i\,
    \frac{\mu}{2}\, \varphi^2 + i\, \frac{g}{4}\, \varphi^4\Bigr\}\, d\varphi\, d\phi} \; .
\end{equation*}

It should be noted, though, that depending on the chosen contour $(\Gamma^0, \Gamma^+,
\Gamma^-)$, the above integral will resemble a Fresnel one and, as such, its computation
is quite simplified.

For the graphs below, we used the following values for the saddle-points,

\begin{equation*}
    \phi_0 \in \Bigl\{0, \pm\hksqrt{-\mu/2\,g}\Bigr\} \; ,
\end{equation*}
where $S'[\phi_0] = 0$. Therefore, the different importance sampling functions are:

\begin{description}
\item[Symmetric Phase] Given by the $\Gamma^0$ contour, this is the one which is regular
  in the limit $g\rightarrow 0$ and accessible via perturbation theory,
  \begin{equation*}
    W_{\epsilon}^{0}[\phi] = \exp\biggl\{-\frac{1}{2}\,
      \Bigl(\frac{\phi}{\epsilon}\Bigr)^2\biggr\} \; ; \quad \phi_0 = 0 \; ;
  \end{equation*}
\item[Solitonic Phase] This phase is given by the linear combination of the
  contours $\Gamma^{+}$ and $\Gamma^{-}$ such that $\mu \geqslant 0$, i.e., this
  represents the soliton solution,
  \begin{equation*}
    W_{\epsilon}^{\mu\geqslant 0}[\phi] = \exp\biggl\{-\frac{1}{2}\,
      \Bigl(\frac{\phi^2 + (\phi_0^{+})^2}{\epsilon^2}\Bigr)\biggr\} + \exp\biggl\{-\frac{1}{2}\,
      \Bigl(\frac{\phi^2 - (\phi_0^{-})^2}{\epsilon^2}\Bigr)\biggr\} \; ;\quad \phi_0^{\pm} = \pm
      i\, \hksqrt{\mu/2\,g \,{}\xspace} \; ; \quad (\mu \geqslant 0)\; ;    
  \end{equation*}
\item[Broken-Symmetric Phase] This one is given by the linear combination of the
  contours $\Gamma^{+}$ and $\Gamma^{-}$ such that $\mu < 0$, i.e., this
  represents the solution usually referred to as broken-symmetric,
  \begin{equation*}
    W_{\epsilon}^{\mu < 0}[\phi] = \exp\biggl\{-\frac{1}{2}\,
      \Bigl(\frac{\phi + \phi_0^{+}}{\epsilon}\Bigr)^2\biggr\} + \exp\biggl\{-\frac{1}{2}\,
      \Bigl(\frac{\phi - \phi_0^{-}}{\epsilon}\Bigr)^2\biggr\} \; ;\quad \phi_0^{\pm} = \pm
      \hksqrt{-\mu/2\,g \,{}\xspace} \; ; \quad (\mu < 0)\; .    
  \end{equation*}
\end{description}

Below, we have the graphs of these importance sampling functions:
\vspace{-2em}
\begin{center}
  \includegraphics[scale=0.55]{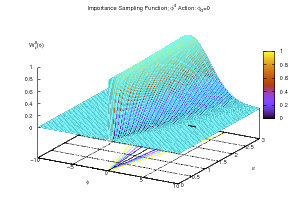}
  \includegraphics[scale=0.55]{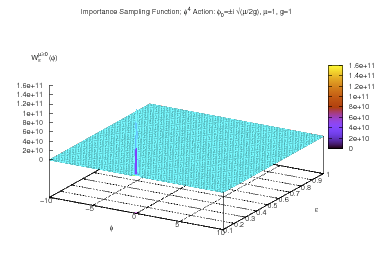}
  \includegraphics[scale=0.55]{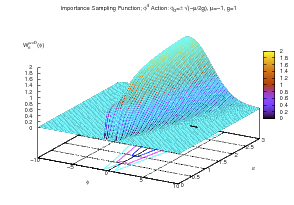}

  \bigskip \bigskip

  {\footnotesize\noindent\textbf{Figure 14:} Plots of the mollified importance
    sampling functions for various values of the field and $\epsilon$: symmetric
  ($\phi_0 = 0$), solitonic ($\mu = 1,\, g = 1$) and broken-symmetric ($\mu =
  -1,\, g = 1$).}
\end{center}

The graphs below show the entropy for the above importance sampling functions, as
discussed in Appendix \ref{sec:entropy}:

\begin{center}
  \hspace{\fill}
  \includegraphics[scale=0.28]{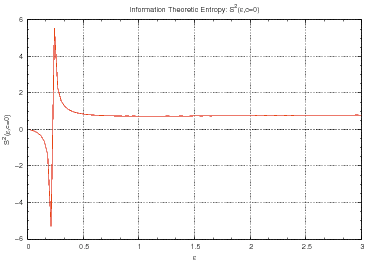}
  \hspace{\fill}
  \includegraphics[scale=0.28]{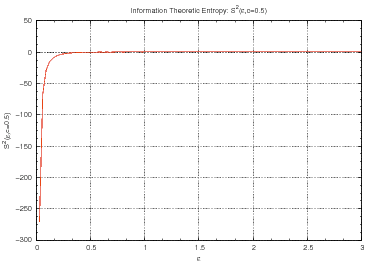}
  \hspace{\fill}
  \includegraphics[scale=0.28]{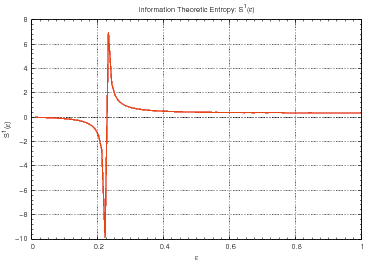}
  \hspace{\fill}

  \bigskip \bigskip

  {\footnotesize\noindent\textbf{Figure 15:} Information theoretical entropy for
  finding the optimal $\epsilon$ in the symmetric, solitonic and
  broken-symmetric cases.}
\end{center}
\subsection{Higher Dimensional}\label{subsubsec:hd}
As for the 4-dimensional results [for a Free Scalar QFT] the news is not
encouraging. As mentioned in section \ref{subsubsec:mis}, there are severe 
efficiency constraints in this implementation of the mollification; in section
\ref{subsubsec:mis} three possible choices of solutions to this technical issue
were given.

It turns out that, from those three solutions, one is non-efficient while the other
two produce analogous results: only the imaginary parts of physically interesting
quantities (generating functional and Green's functions) can be tamed. The real parts
continue to show an oscillatory pattern.

Although this is at least disappointing, the solutions obtained are good enough
to control the imaginary part's oscillations.

It is our hope that changes in algorithms will allow some version of
mollification to work in real world problems.
\acknowledgments
The authors would like to thank S. Corley for discussions and help
with the early stages of this paper. We would also like to thank
J.D. Doll and D. Sabo for useful conversations. Much of the above is an
extension of earlier work done with them. This work is supported in
part by funds provided by the US Department of Energy (\textsf{DoE})
under \textsf{DE-FG02-91ER40688-TaskD}.
\appendix
\section{Convolutions and Smoothing} \label{sec:background}
With the aid of the tools developed below, the smooth approximations of functions can be
done in a mathematically rigorous fashion. This is useful to justify the
statements made in this work, where this technology is applied to the generating
functional of an arbitrary QFT. For proofs of the theorems shown below, see
\cite{pde} \cite{mmmp2}.
\subsection{Mollifiers}\label{subsubsec:mollifiers}
\begin{notat}
  If $U\subset \mathbb{R}^n$ is open, $\partial U$ is its boundary and $\varepsilon > 0$, let
  $U_{\varepsilon} = \{x\in   U\; |\; \mathrm{dist}(x,\partial U) >
  \varepsilon\}$. Further, let $B(0,\varepsilon)$ be the ball centered on $0$ and with
  radius $\varepsilon$.
\end{notat}

\begin{deff}\label{deff:mollifier}
  (Standard mollifier.)

  A \emph{mollifier}, $\eta$, also called an \emph{approximate identity}, is a positive
  $\cont{\infty}\/(\mathbb{R}^n)$ function. The \emph{standard mollifier} is defined in
  the following way:

  \begin{itemize}
    \item Define $\eta\in \cont{\infty}(\mathbb{R}^n)$ to be,
      \begin{equation*}
        \eta(x) =
        \begin{cases}
          C\, \exp\Bigr(\frac{1}{\abs{x}^2 - 1}\Bigr), & \text{if } \abs{x} < 1; \\
          0, & \text{if } \abs{x} \geqslant 1.
        \end{cases}
      \end{equation*}
      The constant $C>0$ selected so that $\int_{\mathbb{R}^n} \eta(x)\, dx = 1$.
    \item $\forall\, \varepsilon >0$, set
      \begin{equation*}
        \eta_{\varepsilon}(x) = \frac{1}{\varepsilon^n}\, \eta(x/\varepsilon) \; .
      \end{equation*}
  \end{itemize}
  $\eta$ is called the \emph{standard mollifier}. The functions $\eta_{\varepsilon}\in
  \cont{\infty}$ satisfy $\int_{\mathbb{R}^n} \eta_{\varepsilon}(x)\, dx = 1$ and
  $\mathrm{supp}(\eta_{\varepsilon}) \subset B(0,\varepsilon)$.
\end{deff}

\begin{deff}\label{deff:mollification}
  (Mollification.) If $f:\, U\to\mathbb{R}$ is locally integrable, define its
  mollification to be,
  \begin{equation*}
    U_{\varepsilon}\ni\; f^{\varepsilon} = \conv{\eta_{\varepsilon}}{f}\; .
  \end{equation*}
  That is, $\forall \; x\in U_{\varepsilon}$,
  \begin{equation}
    \label{eq:moll-conv}
    f^{\varepsilon}(x) = \int_U \eta_{\varepsilon}(x-y)\, f(y)\, dy =
      \int_{B(0,\varepsilon)} \eta_{\varepsilon}(y)\, f(x-y)\, dy \; .
  \end{equation}
\end{deff}
\subsection{Properties of Mollifiers}\label{subsubsec:propmollifiers}
\begin{thm}
  (Properties of mollifiers.)

  \begin{enumerate}
    \item $f^{\varepsilon} \in \cont{\infty}(U_{\varepsilon})$;
    \item $f^{\varepsilon} \to f$, almost everywhere, as $\varepsilon \to
      0$;
    \item If $f\in \cont{}(U)$, then $f^{\varepsilon} \to f$ uniformly on compact
      subsets of $U$; \&
    \item If $1 \leqslant p < \infty$ and $f\in L^p_{\mathrm{loc}}(U)$, then
      $f^{\varepsilon} \to f$ in $L^p_{\mathrm{loc}}(U)$.
  \end{enumerate}
\end{thm}
\section{Entropy Calculations} \label{sec:entropy}
In the cases considered in \ref{subsubsec:ld}, the information theoretic entropy can be
computed analytically. There are 2 types of functions to be considered:

\begin{align*}
  f^1_{\epsilon}(x) &= \exp\biggl\{-\frac{1}{2}\, \biggl(\frac{x +
    c}{\epsilon}\biggr)^{2}\biggr\} \; ;\\
  f^2_{\epsilon}(x) &= \exp\biggl\{-\frac{1}{2}\, \biggl(\frac{x^{2} +
    c}{\epsilon^{2}}\biggr)\biggr\} \; ;
\end{align*}
where $c$ is a constant scalar. The first type above deals with the Airy Function action
for $J < 0$ ($c = \pm \hksqrt{-J}$) and the $\phi^4$ action with $\mu < 0$ ($c = \pm\hksqrt{-\mu/2\,
  g}$), while the second type above handles the Airy Function action for $J \geqslant 0$
($c = \pm J$) and the $\phi^4$ action for $\mu \geqslant 0$ ($c = \pm i\,
\hksqrt{\mu/2\, g}$).

The information theoretic entropy has the general form,

\begin{align*}
  \mathcal{S}^i(\epsilon) &= -\frac{\int f^i_{\epsilon}(x)\,
    \log\bigl(f^i_{\epsilon}(x)\bigr)\, dx}{\int f^i_{\epsilon}(x)\, dx +
    \log\biggl(\int f^i_{\epsilon}(x)\, dx\biggr)} \; .\\
  \intertext{Therefore, we see that the analytical results are given by:}
  \mathcal{S}^1(\epsilon) &= \frac{(\epsilon/2)\, \hksqrt{2\, \pi}}{\hksqrt{2\, \pi}\,
    \epsilon + \log\bigl(\hksqrt{2\, \pi}\, \epsilon\bigr)} \; ;\\
  \mathcal{S}^2(\epsilon,c) &= \frac{e^{-c/2\, \epsilon^2}\,\hksqrt{2\, \pi}\,
    \bigl(\epsilon + c/2\,\epsilon\bigr)}{e^{-c/2\, \epsilon^2}\, \hksqrt{2\, \pi}\,
    \epsilon + \log\bigl(e^{-c/2\, \epsilon^2}\, \hksqrt{2\, \pi}\, \epsilon\bigr)} \; .
\end{align*}

And, thus, we have closed forms for the entropies of the examples given. The graphs below
show these entropies: the leftmost one shows $\mathcal{S}^1(\epsilon)$ while the rightmost
one shows $\mathcal{S}^2(\epsilon,c)$ for a range of $c$ values.

\begin{center}
  \hspace{\fill}
  \scalebox{0.5}{\includegraphics{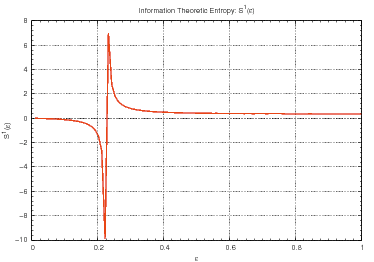}}
  \hspace{\fill}
  \scalebox{0.6}{\includegraphics{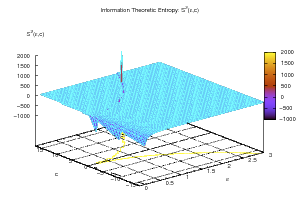}}
  \hspace{\fill}
\end{center}
\section{Parabolic Cylinder Functions} \label{sec:pcf}
The differential equation for the Parabolic Cylinder functions is usually
written as,

\begin{equation*}
  \frac{d² f}{dz²} + (a\, z^2 + b\, z + c)\, f = 0 \; ;
\end{equation*}
where $z \in \mathbb{C}$.

The 0-dimensional $\phi^4$ action is given by $S[\phi] = \mu\, \phi^2/2 + g\,
\phi^4/4$, which yields two Schwinger-Dyson equations:

\begin{align*}
  \text{Perturbative solution:}\qquad & -i\, \delta_{J}\mathcal{Z} - J\,
    \mathcal{Z} = 0 \; ; \\
  \text{Non-perturbative solutions:}\qquad & \delta^{2}_{J}\mathcal{Z} + J\,
    \mathcal{Z} - \beta\, \mathcal{Z} = 0 \; ;
\end{align*}
where $\beta = \mu/g$. The match to the previous form is obtained when $a = 0,\,
b = +1,\, c = -\beta$. Note that there are two non-perturbative solutions, one
being solitonic and the other one being called simply broken-symmetric.

\begin{thebibliography}{999}
  \bibitem{analysis} Elliott H. Lieb, Michael Loss, \emph{Analysis}, GSM, American
    Mathematical Society, 2001.
  \bibitem{pp1} S. B. Fahy, D. R. Hamann, \emph{Positive-projection Monte
      Carlo simulation: A new variational approach to strongly interacting
      fermion systems}, \prl{65}{1990}{3437}.
  \bibitem{pp2} S. B. Fahy, D. R. Hamann, \emph{Diffusive behavior of states
      in the Hubbard-Stratonovitch transformation}, \prb{43}{1991}{765}.
  \bibitem{fd} M. Suzuki, \emph{Fractal decomposition of exponential operators
      with applications to many-body theories and Monte Carlo simulations},
    \pla{146}{1990}{319}.
  \bibitem{bp} P. de Vries, Ph.D. Thesis, University of Amsterdam, 1991,
    unpublished.
  \bibitem{sm} A. Muramatsu, G. Zumbach, X. Zotos, \emph{A Geometrical View
      of the Minus-Sign Problem}, \ijmpc{3}{1992}{185}.
  \bibitem{dtta} A.H. Zemanian, \emph{Distribution Theory and Transform
      Analysis: An Introduction to Generalized Functions, with Applications},
    Dover, 1987.
  \bibitem{stcqp} D. Sabo, J.D. Doll, D.L. Freeman, \emph{Stationary Tempering
      and the Complex Quadrature Problem},
    \newjournal{J. Chem. Phys.}{JCPSA}{116}{2002}{3509}.
  \bibitem{tvbcsde} S. Garcia, Z. Guralnik, G.S. Guralnik, \emph{Theta Vacua and
      Boundary Conditions of the Schwinger-Dyson Equations},
    [\hepth{9612079}]. G.S. Guralnik, Z. Guralnik, \emph{Complexified Path
      Integrals and the Phases of Quantum Field Theory}, [\arXivid{0710.1256}].
  \bibitem{ffisip} M.G.G. Laidlaw, C.M. DeWitt, \emph{Feynman
      Functional Integrals for Systems of Indistinguishable Particles},
    \prd{3}{1971}{1375}.
  \bibitem{nspmc} J.D. Doll, \emph{Notes on the Stationary Phase Monte Carlo
      Method}, unpublished, 1999.
  \bibitem{rtnnm} R. Easther, D.D. Ferrante, G.S. Guralnik and D. Petrov,
    \emph{A review of two novel numerical methods in QFT}, [\heplat{0306038}].
  \bibitem{mmc} D.D. Ferrante, J. Doll, G.S. Guralnik, D. Sabo, \emph{Mollified
      Monte Carlo}, \npps{119}{2003}{965}, [\heplat{0209053}].
  \bibitem{selcrngws} Giovanni Ossola, Alan D. Sokal, \emph{Systematic errors
      due to linear congruential random-number generators with the Swendsen-Wang
      algorithm: A warning}, \pre{70}{2004}{027701}, [\heplat{0403010}].
  \bibitem{mtrng} M. Matsumoto, T. Nishimura, \emph{Mersenne Twister: A
      623-dimensionally equidistributed uniform pseudorandom number generator},
    \newjournal{ACM Trans. on Modeling and Computer
      Simulation}{ATMCS}{8}{1998}{3}.
  \bibitem{ugosumcmci} Tuomas J. Lukka, Janne V. Kujala, \emph{Using Genetic
      Operators to Speed up Markov Chain Monte Carlo Integration},
    \newjournal{Monte Carlo Methods and Appl.}{MCMAP}{8}{2002}{51},
    [\heplat{0403010}].
  \bibitem{Baez} J.C. Baez, \emph{Diffeomorphism invariant generalized
      measures on the space of connections modulo gauge transformations},
    [\hepth{9305045}].
  \bibitem{Ashtekar} A. Ashtekar, J. Lewandowski, \emph{Differential
      geometry on the space of connections via graphs and projective limits},
    \jgp{17}{1995}{191}, [\hepth{9412073}].
  \bibitem{llsp1} L.D. Landau, E.M. Lifshitz, \emph{Statistical Physics}, Part
    1, Butterworth-Heinemann, 1984.
  \bibitem{pde} Lawrence C. Evans, \emph{Partial Differential Equations}, GSM,
    American Mathematical Society, 1998.
  \bibitem{mmmp2} Michael Reed, Barry Simon, \emph{Methods of Modern
      Mathematical Physics, Vol II: Fourier Analysis, Self-Adjointness},
    Academic Press, 1975.
\end{thebibliography}
\end{document}